# Modeling Biphasic, Non-Sigmoidal Dose-Response Relationships: Comparison of Brain-Cousens and Cedergreen Models for a Biochemical Dataset

Venkat D. Abbaraju[1], Tamaraty L. Robinson[2], Brian P. Weiser[2,*]

[1]Virginia Commonwealth University School of Medicine, Richmond, VA  23298 (USA)
[2]Rowan-Virtua School of Translational Biomedical Engineering & Sciences (Rowan University), Department of Molecular Biology, Stratford, NJ  08084 (USA)

*corresponding author: weiser@rowan.edu

## Abstract

Biphasic, non-sigmoidal dose-response relationships are frequently observed in biochemistry and pharmacology, but they are not always analyzed with appropriate statistical methods. Here, we examine curve fitting methods for "hormetic" dose-response relationships where low and high doses of an effector produce opposite responses. We provide the full dataset used for modeling, and we provide the code for analyzing the dataset in SAS using two established mathematical models of hormesis, the Brain-Cousens model and the Cedergreen model. We show how to obtain and interpret curve parameters such as the $ED_{50}$ that arise from modeling, and we discuss how curve parameters might change in a predictable manner when the conditions of the dose-response assay are altered. In addition to modeling the raw dataset that we provide, we also model the dataset after applying common normalization techniques, and we indicate how this affects the parameters that are associated with the fit curves. The Brain-Cousens and Cedergreen models that we used for curve fitting were similarly effective at capturing quantitative information about the biphasic dose-response relationships.

## Introduction

Biphasic, non-sigmoidal dose-response relationships have been observed in many biological systems on the molecular, cellular, and organism levels.[1] The term "hormesis" is used to describe biphasic dose-response relationships where low and high doses produce opposite responses. Hormetic dose-responses are sometimes also called bell-shaped, J-shaped, inverted U-shaped, and other field-specific terms. Hormetic dose-response relationships are not limited to specific types of biological components, drugs, compounds, stimulants, or toxins.[1]

Although hormetic dose-response relationships are common in biochemical assays, they are not always analyzed with appropriate curve fitting techniques that provide—at the least—an accurate $ED_{50}$ (see Methods for parameter descriptions). Additionally, the steepness of the slopes on a biphasic curve are important for understanding the dose ranges that transition the system between high and low responses, much like the Hill coefficient ($n$) of standard sigmoidal dose-responses. Other information including the exact magnitude of a biphasic increase or decrease relative to a control response is lost without proper modeling. The complete characteristics of hormetic dose-responses that are obtained through modeling are essential to accurately predict how a system will respond when dosed and to reliably choose doses that will produce a desired outcome.

It has been proposed that quantitative features of hormetic dose-response relationships are widely conserved regardless of the organism or the biological system used.[1] This will only be clear by comparing of a wide variety of dose-response relationships from different biological systems that are analyzed with identical methods. Here, we analyzed a sample biochemical dataset with mathematical models of hormesis that are commonly used in other fields such as weed science.[2,3] The biochemical assay that was used to collect this data has been described in detail along with the molecular mechanism that underlies the hormetic dose-response.[4–6] The present article focuses on comparing two methods that can be used for curve fitting and the meaning behind the parameters associated with the curves. The methods we describe can be applied to a broader range of molecular systems that exhibit hormesis.

## Methods

### Interpretation and Implementation of Hormetic Dose-Response Models

In simple molecular systems showing hormesis, we refer to the dose variable as the *effector* that induces a response from the system (Figure 1). A second variable is the *substrate* that responds to the effector. The substrate concentration can be changed in dose-response assays, but should remain constant throughout any individual dose-response curve. It is possible for an effector to have more than one substrate in a system, and additional components might be required to produce or observe a response. Every component in a system has the potential to alter the characteristics of a dose-response relationship.

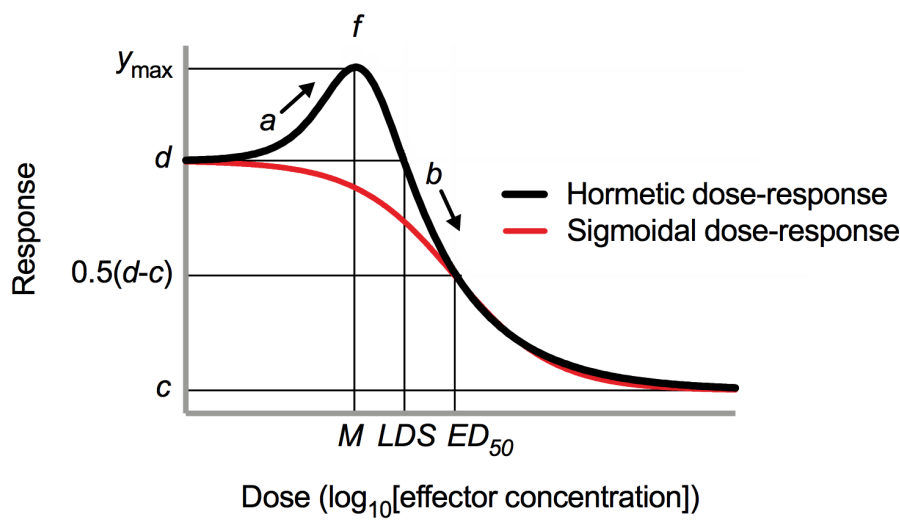

**Figure 1.** Hypothetical sigmoidal dose-response relationship (red trace) and hormetic dose-response relationship (black trace). Refer to the main text for a description of the parameters defined on the curves; note that parameters $a$, $f$, and $b$ do not directly interpolate to the $x$ or $y$ axes, but their approximate meaning was shown here for reference. This figure was adapted from several sources.[2–4,7]

The full dataset used for modeling in this work can be found in the Appendix (Table A1), and its experimental origin has been described.[4] The numerical values of the responses (i.e., the raw data values) were unitless and derived from fluorescence anisotropy-based binding assays;[4–6] their meaning is not discussed here. Data was collected by measuring the response of a substrate to different doses of effector, then in separate experiments, the amount of substrate was changed and the effector was again dosed. Five different concentrations of substrate were used in dose-response assays, and the effector was dosed across the same concentration range for each amount of substrate. The substrate concentrations ranged between 0.25-10 µM, and the effector concentrations ranged between 0.01-50 µM (Table A1).

The data in Table A1 was analyzed with equations that arise from the Brain-Cousens model of hormesis (equations 1-5) and the Cedergreen model of hormesis (equations 6-10).[2,3,8,9] Equations 1 and 6 were the original equations by Brain and Cousens and Cedergreen et al., respectively, which were modifications of a common log-logistic function.[3,9] The remaining equations were parameterizations of equations 1 and 6 that were used to extract additional information from the curves (discussed below).[2,8] Note that our hormetic dataset provided inverted U-shaped curves where the highest doses of effector produced the lowest responses, which directed our description of curve parameters. However, these equations can be used to model hormetic datasets of all shapes.[3]

Brain-Cousens Model

$$y = c + \frac{(d-c) + f*x}{1 + \exp(b*\ln(\frac{x}{e}))} \quad \text{(Eq. 1)}$$

$$y = c + \frac{(d-c) + f*x}{1 + \left(\frac{50}{100-50} + \frac{100}{100-50} * \frac{f*ED50}{d-c}\right)*\exp(b*\ln(x/ED50))} \quad \text{(Eq. 2)}$$

$$y = c + \frac{(d-c) + f*x}{1 + \left(\frac{f*M}{((d-c)*b) - f*M*(1-b)}\right)*\exp\left(b*\ln\left(\frac{x}{M}\right)\right)} \quad \text{(Eq. 3)}$$

$$y = c + \frac{(d-c) + f*x}{1 + \left(\frac{f*LDS}{d-c}\right)*\exp\left(b*\ln\left(\frac{x}{LDS}\right)\right)} \quad \text{(Eq. 4)}$$

$$y_{max} = c + \frac{(d-c) + f*M}{1 + \left(\frac{f*M}{((d-c)*b) - f*M*(1-b)}\right)*\exp\left(b*\ln\left(\frac{M}{M}\right)\right)} \quad \text{(Eq. 5)}$$

Cedergreen Model

$$y = c + \frac{(d-c) + f*\exp(\frac{-1}{x^a})}{1 + \exp(b*\ln(\frac{x}{e}))} \quad \text{(Eq. 6)}$$

$$y = c + \frac{\left(\left(\left(\frac{100-50}{100} - \frac{1}{1+\exp(b*\ln(\frac{ED50}{e}))}\right)^{-1} * \left(\frac{-c+f*\exp(\frac{-1}{ED50^a})}{1+\exp(b*\ln(\frac{ED50}{e}))} + \frac{c*(100-50)}{100}\right)\right) - c\right) + f*\exp(\frac{-1}{x^a})}{1 + \exp(b*\ln(\frac{x}{e}))} \quad \text{(Eq. 7)}$$

$$y = c + \frac{(d-c) + \left(\left(\exp(\frac{-1}{M^a})*(a*M^{-a-1})*(1+\exp(b*\ln(\frac{M}{e}))) - \exp(\frac{-1}{M^a})*\exp(b*\ln(\frac{M}{e}))*\frac{b}{M}\right)^{-1} * \left((d-c)*\exp(b*\ln(\frac{M}{e}))*\frac{b}{M}\right)\right)*\exp(\frac{-1}{x^a})}{1+\exp(b*\ln(\frac{x}{e}))} \quad \text{(Eq. 8)}$$

$$y = c + \frac{\left(\left(\left(1 - \frac{1}{1+\exp(b*\ln(\frac{LDS}{e}))}\right)^{-1} * \left(\frac{-c+f*\exp(\frac{-1}{LDS^a})}{1+\exp(b*\ln(\frac{LDS}{e}))} + c\right)\right) - c\right) + f*\exp(\frac{-1}{x^a})}{1+\exp(b*\ln(\frac{x}{e}))} \quad \text{(Eq. 9)}$$

$$y_{max} = c + \frac{(d-c) + \left(\left(\exp(\frac{-1}{M^a})*(a*M^{-a-1})*(1+\exp(b*\ln(\frac{M}{e}))) - \exp(\frac{-1}{M^a})*\exp(b*\ln(\frac{M}{e}))*\frac{b}{M}\right)^{-1} * \left((d-c)*\exp(b*\ln(\frac{M}{e}))*\frac{b}{M}\right)\right)*\exp(\frac{-1}{M^a})}{1+\exp(b*\ln(\frac{M}{e}))} \quad \text{(Eq. 10)}$$

A brief description of parameters found in the Brain-Cousens and Cedergreen models of hormesis are as follows:

   *a*. Parameter *a* is the only parameter that is specific to the Cedergreen model, and is not found in the Brain-Cousens model. Parameter *a* was reported to control the rate of hormetic increase prior to the hormetic peak.[3,10]

   *b*. Parameter *b* controls the steepness of the descending part of the curve towards the lower asymptote after the hormetic peak. Parameter *b* is analogous to the commonly used Hill coefficient *n*; however in the equations used here, *b* is positive in the downward sloping responses (i.e., *b* and *n* have conventionally opposite signs). In practice, *b* controls the dose distance between (for example) the *LDS* and $ED_{50}$.

   *c*. Parameter *c* is the *y* response value at the lower asymptote and is the theoretical response at infinite doses of effector (Figure 1).

*d*. Parameter *d* is the *y* response value at the upper asymptote and is the theoretical control response in the absence of effector (Figure 1).

*e*. Parameter *e* in hormetic modeling provides a lower bound on the $ED_{50}$ but has no straightforward biological meaning.[3,8,9] Parameter *e* and the $ED_{50}$ are the same in the absence of hormesis.[3,10]

*f*. Parameter *f* is the hormesis parameter. Parameter *f* should equal 0 in the absence of hormesis which reverts the hormetic equations back to a standard sigmoidal log-logistic function (in other words, set *f* = 0 in equations 1 and 6 to obtain an equation for a standard sigmoidal dose-response relationship). In practice, a value for *f* is always determined during modeling even if it is infinitely close to 0. A hormetic effect is often confirmed with some statistical confidence when the 95% confidence interval of parameter *f* does not overlap with the value 0.[2,8] As cautioned,[7,11] the actual magnitude of *f* is not always directly related to the size of the hormetic increase.

$ED_{50}$. The $ED_{50}$ is the effective dose *x* that reduces the response *y* to the halfway point between *d* and *c* (Figure 1). The $ED_{50}$ on a standard sigmoidal dose-response curve is often called the $EC_{50}$ (half maximal effective concentration) or $IC_{50}$ (half maximal inhibitory concentration).

*M*. Parameter *M* is the dose *x* that provides the maximum stimulatory response *y* (i.e., $y_{max}$).

*LDS*. The *LDS* (limiting dose for stimulation) is the highest dose *x* where the hormetic increase vanishes and the response *y* returns to the value of *d* or the upper asymptote.

$y_{max}$. The $y_{max}$ is the maximum stimulatory response *y* that occurs at the hormetic peak.

$y_{max}\%$. The $y_{max}\%$ is calculated as the percent change between the control response *d* and $y_{max}$. [$y_{max}\% = (y_{max} / d) * 100\%$]

For data modeling, the NLMIXED procedure of SAS software was used to fit response values *y* as a nonlinear biphasic function of dose *x*. All calculations were performed with SAS Studio OnDemand for Academics webserver. The code for our calculations in SAS can be found in the Appendix. For the Brain-Cousens analyses, parameters *b*, *d*, *e*, and *f* were determined using equation 1, whereas $ED_{50}$, *M*, and *LDS* were calculated using the relevant parameterization (equations 2, 3, and 4). Parameter $y_{max}$ was calculated using equation 5 with *x* = *M*. For the Cedergreen model analyses, parameters *a*, *b*, *d*, *e*, and *f* were determined using equation 6, while $ED_{50}$, *M*, *LDS* and $y_{max}$ were determined using equations 7, 8, 9, and 10. When calculating values for parameters $ED_{50}$, *M*, *LDS*, and $y_{max}$, the values for *a*, *b*, *d*, *e*, and *f* were fixed to the values that were obtained with equations 1 and 6. It is statistically preferred to solve for all parameters independently without introducing these fixed values into the equations, but in practice, we found equations 1 and 6 to be significantly more robust in their ability to model the datasets compared to the parameterizations (with the exception of equation 2, which we also found highly reliable). Therefore, fixing the values improved modeling efficiency, and as indicated below, the initial parameters determined with equations 1 and 6 were used to gauge modeling quality (i.e., goodness-of-fit). Furthermore, parameter *c* was fixed in all of our analyses. Generally speaking, fixing parameter *c* to a specific value while using any of equations 1-10 is expected to occur more frequently than fixing other parameters as a consequence of technical limitations that prevent data collection at high doses such as limited effector solubility at high concentrations or effector availability. With our raw dataset, a value of 0.0572 for parameter *c* could readily be obtained using the Brain-Cousens equations by analyzing the 0.25 µM substrate data without constraints, or by analyzing the 0.25 µM substrate data with a standard sigmoidal dose-response equation.[4] At its simplest, 0.0572 was also the average of the data values for the 0.25 µM substrate data at a dose of 50 µM effector (Table A1), which rested on the lower asymptote. In the experimental paradigm, the response of the substrate to an infinite dose of effector was theoretically the same regardless of the substrate concentration which justified the universal constraint on *c*.

The modeling procedure required that "starting values" were provided for each parameter in the software code; the importance of the starting values in facilitating the algorithms' convergence to reported parameter values has been extensively discussed.[10] A general recommendation is to deduce starting values for each parameter after visualizing the plotted datasets on a graph.[8,10] In all analyses, the starting value for *d* was chosen as the approximate value that was expected for the upper asymptote when the 0.25 µM substrate dataset was plotted and visualized on a graph. Starting values for parameters *a*, *b*, *e*, and *f* were determined empirically by testing sets of values that resulted in convergence of the estimation algorithm.[8] Note that the starting values in some analyses required more trial and error than others to achieve satisfactory fits for the curves. Our starting values are reported in the Appendix (Table A2).

To draw the curves that were fit to our datasets in the Figure panels, we solved for *y* responses to hypothetical *x* values using equations 1 and 6. Equations 1 and 6 were also used to calculate $R^2$ values that compared goodness-of-fit for the hormetic models. $R^2$ and adjusted $R^2$ values ($\overline{R^2}$) were calculated using Posit Cloud webserver (formerly RStudio Cloud) using equations 11 and 12, respectively

$$R^2 = 1 - \frac{\sum_{i=1}^{n}(y_i - \hat{y}_i)^2}{\sum_{i=1}^{n}(y_i - \bar{y})^2} \qquad \text{(Eq. 11)}$$

$$\overline{R^2} = 1 - \frac{(1-R^2)(n-1)}{n-k-1} \qquad \text{(Eq. 12)}$$

where $y_i$ was the experimental response value, $\hat{y}_i$ was the predicted response value based on the fit curve, $\bar{y}$ was the mean of the experimental response values, *n* was the total number of data points, and *k* was the number of independent variables. $\overline{R^2}$ was calculated because $R^2$ is highly dependent on the size of the experimental dataset, and $\overline{R^2}$ adjusts for the population size and number of independent variables.[12,13] Note that $R^2$ and $\overline{R^2}$ are generally not recommended measures of goodness-of-fit for non-linear models,[2,14,15] but they remain familiar metrics that represent how much of the change in the response variable was explained by changes in the dose variable. Thus, we justified using these metrics to compare the goodness-of-fit for two models on the same dataset. The RStudio code can be found in the Appendix.

## Results and Discussion

**Fitting Curves to the Raw Hormetic Dataset with Brain-Cousens and Cedergreen Models**

The dose-response data in Table A1 was modeled with Brain-Cousens and Cedergreen equations to yield the parameters shown in Table 1. With the exception of parameter *e*, we previously reported parameters for this raw dataset using the Brain-Cousens equations;[4] the analyses discussed here were independently reproduced. The Brain-Cousens model provided well-fit curves for each dose-response relationship from assays with different substrate concentrations (Figure 2A). This was clear both qualitatively (visually) and quantitatively based on $R^2$ values. For the Brain-Cousens analysis of the 0.25 µM substrate data, the hormesis parameter *f* was not considered statistically different from 0 because its 95% confidence interval overlapped with 0. This dose-response relationship was therefore not significantly hormetic,[2,8] and it qualitatively resembled a standard monotonic sigmoidal dose-response (Figure 2A). Additionally, the Brain-Cousens equations could satisfactorily model the data from experiments that used 10 µM substrate, even though we report some parameters with caution because the descending part of the curve was not adequately represented by experimental data points (i.e., parameters *b*, $ED_{50}$, and *LDS* cannot be certain).

The Cedergreen model also provided well-fit curves for the dose-response assays that used 0.25-3 µM of substrate (Figure 2B). The $R^2$ values for these curves were quantitatively similar to the $R^2$ values determined with the Brain-Cousens model (Table 1). The Cedergreen analyses also suggested that $f$ was not different from 0 for the 0.25 µM substrate data because the 95% confidence interval of the value overlapped with 0. We concluded that the Brain-Cousens and Cedergreen equations were similarly effective at modeling the data from assays that used 0.25-3 µM of substrate. In contrast, we were unable to obtain a reasonable curve for the 10 µM substrate data using the Cedergreen equations. The inability of the Cedergreen equations to model the 10 µM substrate data reflects a higher requirement for stringency in the dataset compared to the Brain-Cousens equations, which could model the data despite there being incomplete inhibition of the response at the highest doses of effector (Figure 2 and Table A1). The Brain-Cousens model likely has lower demands on the quality of the data because it contains one less parameter than the Cedergreen model.

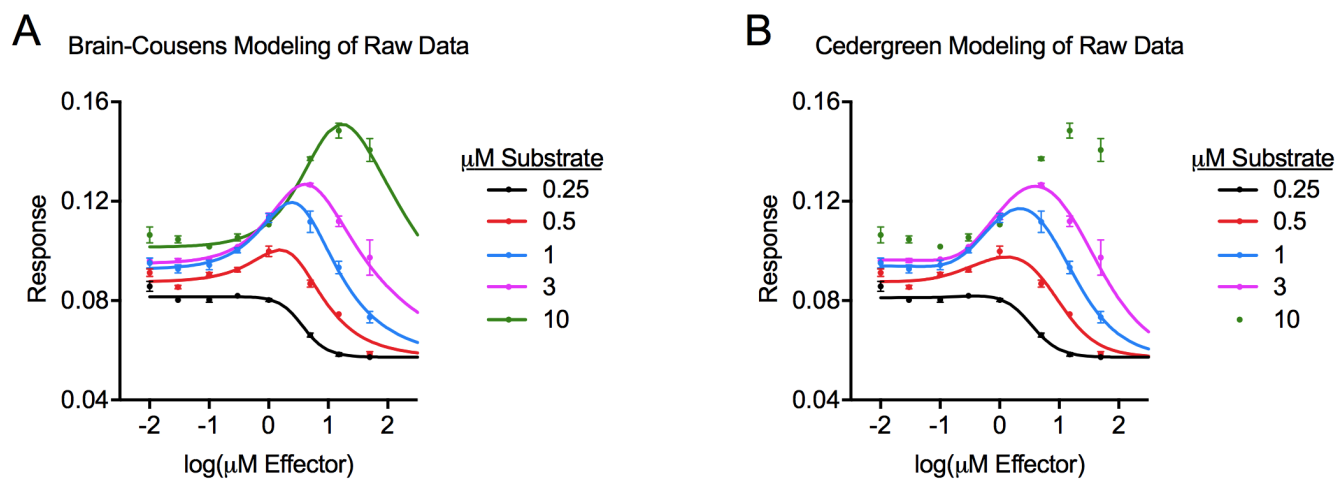

**Figure 2.** Dose-response curves generated from modeling the raw dataset (Table A1) with the Brain-Cousens model (equation 1) (panel A) or the Cedergreen model (equation 6) (panel B).

**Table 1. Dose-Response Curve Parameters and $R^2$ from Modeling the Raw Dataset in Table A1**

|  | Brain-Cousens Model | | | | | Cedergreen Model | | | | |
|---|---|---|---|---|---|---|---|---|---|---|
|  | Substrate Concentration (µM) | | | | | Substrate Concentration (µM) | | | | |
| Parameter | 0.25 | 0.5 | 1 | 3 | 10 | 0.25 | 0.5 | 1 | 3 | 10 |
| $a$ | n/a | n/a | n/a | n/a | n/a | 0.5076 | 0.4510 | 0.6524 | 0.7093 | ND |
| $b$ | 2.1686 | 1.7045 | 1.5779 | 1.3850 | 1.3115 | 1.8270 | 1.2639 | 0.9496 | 0.8739 | ND |
| $c$ | 0.0572[a] | 0.0572[a] | 0.0572[a] | 0.0572[a] | 0.0572[a] | 0.0572[a] | 0.0572[a] | 0.0572[a] | 0.0572[a] | 0.0572[a] |
| $d$ | 0.0815 | 0.0876 | 0.0927 | 0.0949 | 0.1015 | 0.0812 | 0.0877 | 0.0940 | 0.0964 | ND |
| $e$ | 3.5932 | 2.5752 | 2.9472 | 3.6165 | 11.2869 | 3.2469 | 7.2980 | 8.5212 | 26.3103 | ND |
| $f$ | 0.0005[b] | 0.0200 | 0.0300 | 0.0280 | 0.0124 | 0.0057[b] | 0.0352 | 0.0722 | 0.0624 | ND |
| $ED_{50}$ | 3.8469 | 15.6195 | 48.5266 | 286.10 | 4143.85 | 3.7259 | 16.1828 | 42.5065 | 131.87 | ND |
| $M$ | 0.2008 | 1.5522 | 2.4486 | 4.1017 | 16.8549 | 0.3890 | 1.3054 | 2.1721 | 3.9471 | ND |
| $LDS$ | 0.3901 | 5.4594 | 14.3212 | 46.9853 | 447.11 | 0.8044 | 5.6977 | 14.4126 | 41.3476 | ND |
| $y_{max}$ | 0.0816 | 0.1004 | 0.1196 | 0.1269 | 0.1510 | 0.0818 | 0.0976 | 0.1172 | 0.1260 | ND |
| $y_{max}\%$ | 100.07 | 114.67 | 129.05 | 133.63 | 148.75 | 100.77 | 111.30 | 124.64 | 130.80 | ND |
| $R^2$ | 0.9560 | 0.9260 | 0.8003 | 0.8497 | 0.9586 | 0.9538 | 0.9307 | 0.8098 | 0.8324 | ND |
| $R^2_{adjusted}$ | 0.9539 | 0.9225 | 0.7957 | 0.8463 | 0.9565 | 0.9516 | 0.9274 | 0.8055 | 0.8286 | ND |

[a] parameter $c$ was fixed
[b] $f$ was not significantly different from zero based on the 95% confidence interval
Abbreviations: ND, not determined; n/a, not applicable

**Fitting Curves to the Hormetic Dataset after Subtracting Baseline**

As further discussed in the Methods, the lower asymptotes (parameter $c$) for the dose-response relationships from the raw dataset were theoretically identical for all of the assays regardless of the substrate concentration used, and this value was non-zero (0.0572). However, the baseline measurement of 0.0572 represented a complete absence of a response. It could therefore be argued that subtracting this baseline measurement from all of the data values before modeling would more accurately describe the magnitude of the hormetic effect relative to the control (no effector) assay condition. This occurs because baseline subtraction reduces the numerical value of $d$, which is critical for $y_{max}$% calculation ($y_{max}$% = ($y_{max}$ / $d$) * 100%), even though baseline subtraction has no effect on the range of the data values ($d$-$c$). To illustrate this, we modeled the dataset in Table A1 as before using Brain-Cousens and Cedergreen equations, but first we subtracted the baseline signal (0.0572) from all of the data values (Figures 3A and 3B). Subtracting 0.0572 to make the lower asymptote 0 (as if no response had occurred) had the following expected effects on the parameters compared to their values from modeling the raw dataset: parameters $a$, $b$, $e$, $f$, $ED_{50}$, $M$, and $LDS$ were essentially unchanged, and parameters $c$, $d$, and $y_{max}$ were reduced by ~0.0572 (Table 2). $R^2$ values indicated that the modeling was similarly effective using raw data and baseline subtracted data (Table 2). We still could not model the 10 µM substrate data using the Cedergreen equations.

As mentioned above, a notable effect of subtracting the baseline signal was observed on the $y_{max}$% values. Compared to the $y_{max}$% values from modeling the raw dataset (Table 1), baseline subtraction prior to modeling with the Brain-Cousens equations increased the $y_{max}$% by 24%-42% when the substrate concentration was 0.5-10 µM (similar increases were determined with Cedergreen equations) (Table 2). This was a cautionary example where modeling raw data values and processed data provided different parameters, but data processing could be essential for proper determination of hormesis magnitude.

A

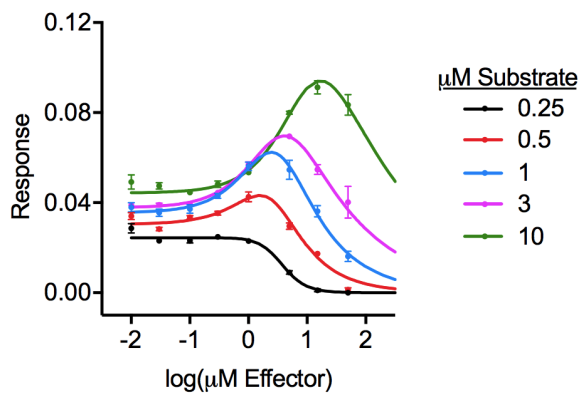

B

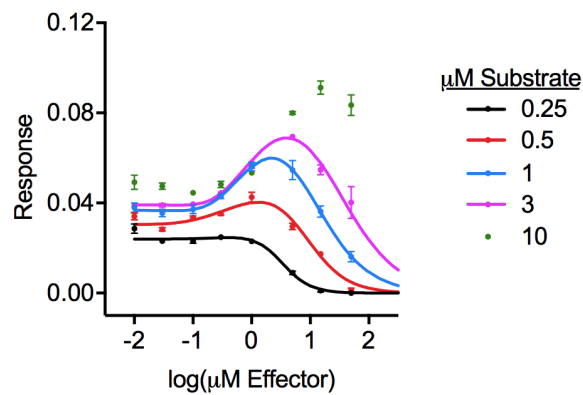

**Figure 3.** Dose-response curves generated from modeling the baseline subtracted dataset with the Brain-Cousens model (equation 1) (panel A) or the Cedergreen model (equation 6) (panel B).

Table 2. Dose-Response Curve Parameters and $R^2$ for Modeling the Dataset with Baseline (0.0572) Subtracted

| | Brain-Cousens Model | | | | | Cedergreen Model | | | | |
|---|---|---|---|---|---|---|---|---|---|---|
| | Substrate Concentration (µM) | | | | | Substrate Concentration (µM) | | | | |
| Parameter | 0.25 | 0.5 | 1 | 3 | 10 | 0.25 | 0.5 | 1 | 3 | 10 |
| $a$ | n/a | n/a | n/a | n/a | n/a | 0.5076 | 0.4510 | 0.6524 | 0.7093 | ND |
| $b$ | 2.1686 | 1.7045 | 1.5779 | 1.3850 | 1.3115 | 1.8270 | 1.2639 | 0.9496 | 0.8739 | ND |
| $c$ | 0[a] | 0[a] | 0[a] | 0[a] | 0[a] | 0[a] | 0[a] | 0[a] | 0[a] | 0[a] |
| $d$ | 0.0243 | 0.0304 | 0.0355 | 0.0377 | 0.0443 | 0.0240 | 0.0305 | 0.0368 | 0.0392 | ND |
| $e$ | 3.5932 | 2.5752 | 2.9472 | 3.6165 | 11.2969 | 3.2470 | 7.2980 | 8.5215 | 26.3103 | ND |
| $f$ | 0.0005[b] | 0.0200 | 0.0300 | 0.0280 | 0.0124 | 0.0057[b] | 0.0352 | 0.0722 | 0.0624 | ND |
| $ED_{50}$ | 3.8469 | 15.6195 | 48.5266 | 286.10 | 4163.61 | 3.7261 | 16.1828 | 42.5082 | 131.87 | ND |
| $M$ | 0.2008 | 1.5522 | 2.4486 | 4.1017 | 16.8752 | 0.3890 | 1.3054 | 2.1722 | 3.9471 | ND |
| $LDS$ | 0.3901 | 5.4594 | 14.3212 | 46.9853 | 449.24 | 0.8044 | 5.6977 | 14.4132 | 41.3476 | ND |
| $y_{max}$ | 0.0244 | 0.0432 | 0.0624 | 0.0697 | 0.0938 | 0.0246 | 0.0404 | 0.0600 | 0.0688 | ND |
| $y_{max}\%$ | 100.22 | 142.30 | 175.88 | 184.61 | 211.93 | 102.61 | 132.51 | 162.94 | 175.81 | ND |
| $R^2$ | 0.9562 | 0.9258 | 0.8003 | 0.8498 | 0.9584 | 0.9538 | 0.9306 | 0.8098 | 0.8324 | ND |
| $R^2_{adjusted}$ | 0.9541 | 0.9223 | 0.7957 | 0.8464 | 0.9563 | 0.9516 | 0.9273 | 0.8055 | 0.8286 | ND |

[a] parameter $c$ was fixed
[b] $f$ is not significantly different from zero based on the 95% confidence interval
Abbreviations: ND, not determined; n/a, not applicable

**Fitting Curves to the Hormetic Dataset after Normalization and Scaling**

Prior to modeling, we normalized and scaled the data values using equation 13

$$y_{normalized\ and\ scaled} = \left(\frac{y_{raw}-c}{d-c}\right) * 100 \tag{Eq. 13}$$

where $y_{raw}$ was the response from the original dataset (Table A1), $c$ was the baseline signal (0.0572), and $d$ was the control response (upper asymptote; Table 1) when the substrate concentration was the same as it was for the $y_{raw}$ value. This transformation subtracted baseline and scaled each dose-response such that the upper asymptote ($d$) was 100 and the lower asymptote ($c$) was 0. This presented the dose-response data as if "% of control" was on the $y$ axis where 100 was the control (no effector) response, and 0 was complete inhibition of the response (Figure 4). This $y$ axis transformation is common when reporting normalized dose-response data in biochemistry and pharmacology, and the multiplier of 100 can be omitted from equation 13 to present data as a "fraction of control" with a response range of 0 to 1.[16] In some cases, normalizing dose-response relationships to a control (no effector) response can remove experimental variability associated with the control level (parameter $d$) without changing values for other parameters such as $ED_{50}$, $b$, $M$, and $LDS$ which are often scale-independent parameters.[16][†]

As observed before, Brain-Cousens and Cedergreen equations were similarly effective at modeling the dataset after normalizing and scaling the values (Figure 4A, Figure 4B, and $R^2$ values in Table 3). During modeling, parameter $c$ and parameter $d$ were fixed to 0 and 100, respectively (Table 3). Because equation 13 subtracted baseline, other parameters in Table 3 should be compared to Table 2. Parameters $a$, $b$, $e$, $ED_{50}$, $M$, $LDS$, and $y_{max}\%$ were essentially unchanged, and $y_{max}$ converted from a value related to experimental measurement (Table 2) to become equivalent to the value of the $y_{max}$% (Table 3). Finally, the numerical value of parameter $f$ increased ~2200-fold to ~4100-fold from Table 2 to Table 3 depending on the substrate concentration and the model used. Parameter $f$ scales with the $y$ response values, and to illustrate this, the $f/d$ and $f/y_{max}$ ratios for individual dose-response conditions were unchanged (Table 4).

---

[†] In this work, the values for parameter $d$ used in equation 13 were determined from modeling and are shown in Table 1. Because Brain-Cousens and Cedergreen modeling determined slightly different values for $d$, the normalized and scaled datasets used for the different models in Figures 4A and 4B were different from each other with data values differing by as much as 3.8%. The fact that we could not determine parameter $d$ for Cedergreen modeling of the 10 µM substrate data is why we could not normalize and scale that data, leading to its absence in Figure 4B and its status as *not applicable* in Table 3. Ordinarily, an experimenter may be inclined to choose a control (no effector) response as parameter $d$ to use with equation 13 as opposed to modeling the raw data first to derive $d$.

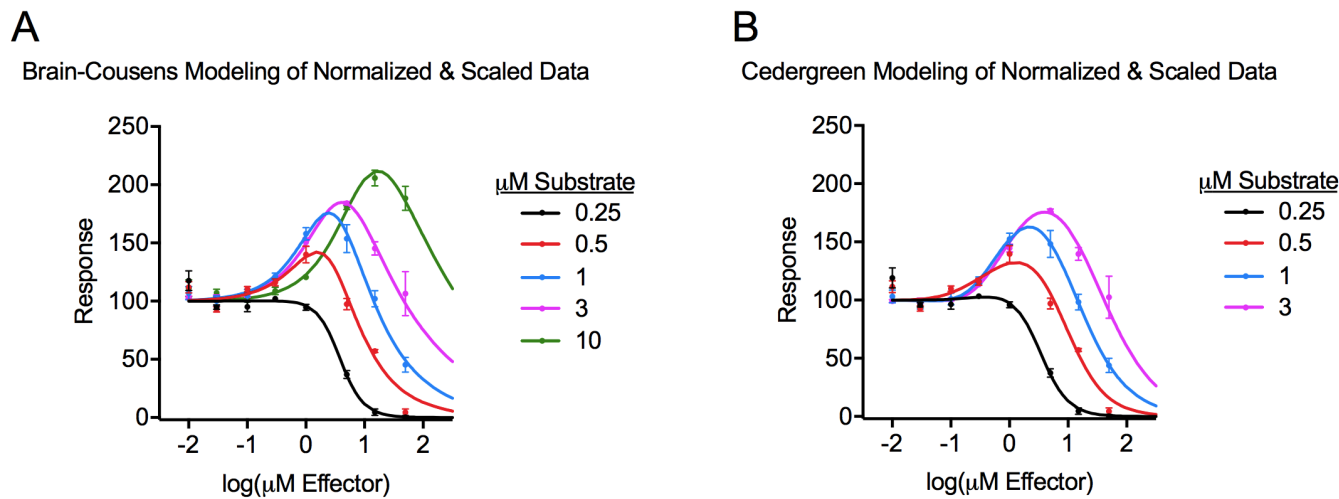

**Figure 4.** Dose-response curves generated from modeling the normalized and scaled dataset with the Brain-Cousens model (equation 1) (panel A) or the Cedergreen model (equation 6) (panel B).

Table 3. Dose-Response Curve Parameters and $R^2$ for Modeling the Dataset after Normalization and Scaling

| | Brain-Cousens Model | | | | | Cedergreen Model | | | | |
| | Substrate Concentration (μM) | | | | | Substrate Concentration (μM) | | | | |
| Parameter | 0.25 | 0.5 | 1 | 3 | 10 | 0.25 | 0.5 | 1 | 3 | 10 |
|---|---|---|---|---|---|---|---|---|---|---|
| $a$ | n/a | n/a | n/a | n/a | n/a | 0.5087 | 0.4525 | 0.6523 | 0.7118 | n/a |
| $b$ | 2.1953 | 1.7047 | 1.5782 | 1.3845 | 1.3130 | 1.8295 | 1.2634 | 0.9497 | 0.8706 | n/a |
| $c$ | 0[a] | 0[a] | 0[a] | 0[a] | 0[a] | 0[a] | 0[a] | 0[a] | 0[a] | n/a |
| $d$ | 100[b] | 100[b] | 100[b] | 100[b] | 100[b] | 100[b] | 100[b] | 100[b] | 100[b] | n/a |
| $e$ | 3.6532 | 2.5803 | 2.9518 | 3.6051 | 11.3454 | 3.2534 | 7.2927 | 8.5220 | 26.2519 | n/a |
| $f$ | 1.7762[c] | 65.6754 | 84.3791 | 74.5299 | 27.7673 | 23.4966[c] | 115.15 | 196.30 | 159.29 | n/a |
| $ED_{50}$ | 3.8745 | 15.6050 | 48.4562 | 287.64 | 4067.54 | 3.7651 | 16.1686 | 42.5023 | 132.39 | n/a |
| $M$ | 0.1917 | 1.5531 | 2.4506 | 4.0958 | 16.8928 | 0.3879 | 1.3056 | 2.1720 | 3.9385 | n/a |
| $LDS$ | 0.3707 | 5.4542 | 14.3082 | 47.1306 | 443.57 | 0.7988 | 5.6877 | 14.4126 | 41.3385 | n/a |
| $y_{max}$ | 100.19 | 142.17 | 175.76 | 184.78 | 211.82 | 102.56 | 132.39 | 162.94 | 175.60 | n/a |
| $y_{max}$% | 100.19 | 142.17 | 175.76 | 184.78 | 211.82 | 102.56 | 132.39 | 162.94 | 175.60 | n/a |
| $R^2$ | 0.9560 | 0.9259 | 0.8003 | 0.8499 | 0.9585 | 0.9538 | 0.9307 | 0.8098 | 0.8328 | n/a |
| $R^2_{adjusted}$ | 0.9539 | 0.9224 | 0.7957 | 0.8464 | 0.9565 | 0.9516 | 0.9274 | 0.8055 | 0.8290 | n/a |

[a] parameter $c$ was fixed
[b] parameter $d$ was fixed
[c] $f$ is not significantly different from zero based on the 95% confidence interval
Abbreviations: ND, not determined; n/a, not applicable

Table 4. Comparison of Parameter Ratios ($f/d$ and $f/y_{max}$) from Baseline Subtracted or Normalized and Scaled Datasets

| | Brain-Cousens Model | | | | Cedergreen Model | | |
|---|---|---|---|---|---|---|---|
| | Substrate Concentration (μM) | | | | Substrate Concentration (μM) | | |
| Parameter Ratio | 0.5 | 1 | 3 | 10 | 0.5 | 1 | 3 |
| $f/d$ (Baseline Subtracted)[a] | 0.6593 | 0.8451 | 0.7421 | 0.2799 | 1.1543 | 1.9620 | 1.5918 |
| $f/d$ (Normalized and Scaled)[b] | 0.6568 | 0.8438 | 0.7453 | 0.2777 | 1.1515 | 1.9630 | 1.5929 |
| $f/y_{max}$ (Baseline Subtracted)[a] | 0.4633 | 0.4808 | 0.4017 | 0.1322 | 0.8710 | 1.2033 | 0.9070 |
| $f/y_{max}$ (Normalized and Scaled)[b] | 0.4619 | 0.4801 | 0.4033 | 0.1311 | 0.8698 | 1.2047 | 0.9071 |

[a]parameter values from Table 2
[b]parameter values from Table 3

**Parameter Characteristics Across Dose-Response Relationships and Comparison to Published Trends**

Generalizable features of hormetic dose-response curves have been reported from analyzing dozens of hormetic datasets that derived from whole organism studies with plants.[7,11,17] We relate these findings to our curves which were produced with the same mathematical models and discuss how parameter values change as a function of substrate concentration. This discussion is restricted to parameters in Table 2, which are graphically represented in Figure 3, and which accurately describe relative response magnitudes to effector doses. Note that the values in Table 1 from Brain-Cousens modeling were also discussed previously,[4] and normalization of the dataset artificially set some parameters to specific values in Table 3, thereby modifying their trends. Finally, the 0.25 μM substrate dataset was not hormetic; in the discussion below, parameters $a$, $f$, $LDS$, $M$, $y_{max}$, and $y_{max}\%$ from the 0.25 μM substrate data do not have any meaning and were therefore excluded from the analysis. Additionally, we excluded parameters $b$, $ED_{50}$, and $LDS$ from the Brain-Cousens modeling of 10 μM substrate assays because of their uncertainty, as the dataset lacked experimental measurements in that part of the curve.

The clearest trend from hormetic dose-response modeling was the strong correlation between $d$ and $y_{max}$ (Figure 5A).[7,11,17] Experimental conditions that produced a higher control response ($d$) in the absence of effector were capable of producing a higher stimulatory response ($y_{max}$) in the presence of effector.[7,11,17] These parameters are real values that can be experimentally measured and are linked with the relationship $d \leq y_{max}$. In contrast, the correlation between $d$ and $y_{max}\%$ was not as strong because, as the control response increases, the system does not necessarily have the ability to produce a proportionately higher hormetic increase—even though this was observed in our dataset. Generally, parameters that trended in the same direction as a function of substrate concentration correlated with each other (in our dataset, $a$, $d$, $ED_{50}$, $M$, $LDS$, $y_{max}$, and $y_{max}\%$ consistently increased as the substrate concentration increased).

It was reported that the actual size of parameters $a$ and $f$, which are specific to the hormetic equations as opposed to sigmoidal curves, are not directly related to the magnitude of hormesis.[7,11] Rather, in plant studies with pronounced hormetic effects, there tended to be low values for $a$ and high values for $f$, and therefore the $f/a$ ratio was more predictive for high values of $y_{max}$.[10,11] Our limited analysis contrasted this conclusion as $a$ consistently increased with greater hormetic effect (Figure 5B), $f$ had no consistent relationship to $y_{max}$ for either the Brain-Cousens or Cedergreen models (Figure 5C), and the $f/a$ ratio was not reliably associated with the magnitude of hormesis (Figure 5D). In prior studies, values for parameter $a$ were in the ~0.1 to 0.7 range,[10,18] which was consistent with the values for $a$ that we determined (0.45 to 0.71).

Finally, parameter $e$ provides a lower bound on the $ED_{50}$ in hormetic curves, but the question of "how much lower" has been posited.[10] In 89 curves from plant experiments, parameter $e$ was on average 1.7-fold lower than the $ED_{50}$, but there was some variability (1.1 to 4.8 fold lower).[10] For our modeling with the Cedergreen equations, parameter $e$ was 2.2 to 5.0-fold lower than the $ED_{50}$. In contrast, parameter $e$ was 6 to 369-fold lower than the $ED_{50}$ in Brain-Cousens modeling, and parameter $e$ increased consistently when the substrate concentration, $ED_{50}$, and the degree of hormesis also increased (Table 2). Even though parameter $e$ has ambiguous significance in a biological context, its potential values are useful to consider because of its presence in equations 1 and 6 that were robustly used for modeling.

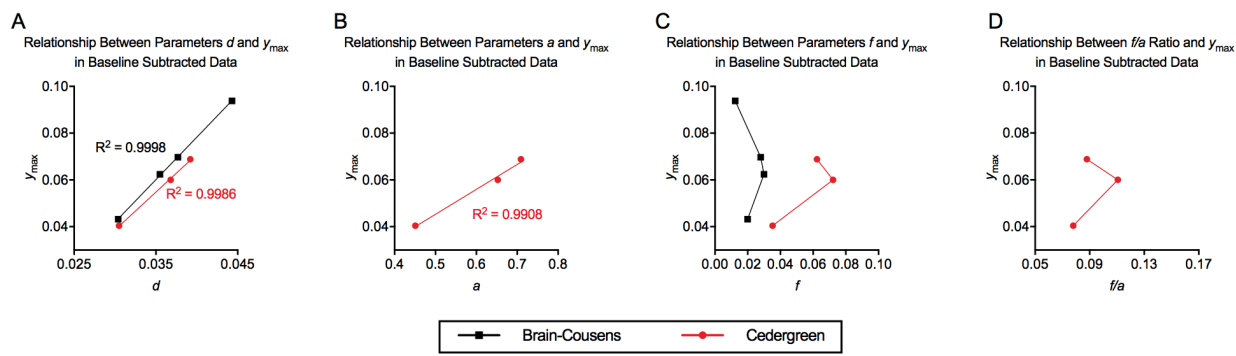

**Figure 5.** (A) Relationship between parameters $d$ and $y_{max}$, (B) Relationship between parameters $a$ and $y_{max}$, (C) Relationship between parameters $f$ and $y_{max}$, and (D) Relationship between $f/a$ ratio and $y_{max}$. Values for the parameters used in this figure are in Table 2.

## Conclusions

This study analyzed a biphasic, biochemical dataset with two mathematical models of hormesis: the Brain-Cousens model and the Cedergreen model. Both models were effective at quantitatively describing the hormetic dose-responses that were collected under different experimental conditions and were processed with different normalization techniques. The comprehensive analyses performed here, along with the review of associated equations, hormetic curve parameters, and software code, provides a framework for analyzing additional hormetic datasets of biochemical and molecular origin. Future investigation will reveal the extent to which hormetic dose-responses from molecular systems quantitatively resemble hormetic dose-responses from cells and organisms.

## Acknowledgements

This work was supported by National Institutes of Health grant R01GM135152. The authors sincerely thank PD Dr. Regina Belz for critical reading of the manuscript and for providing initial code used for hormetic modeling.

# Appendix

Table A1. Raw Dataset Used for Dose-Response Modeling[a]

| μM Effector[b] | 0.25 μM Substrate[c] | | | | 0.5 μM Substrate | | | | 1 μM Substrate | | | | 3 μM Substrate | | | | 10 μM Substrate | | | |
|---|---|---|---|---|---|---|---|---|---|---|---|---|---|---|---|---|---|---|---|---|
| 0.01 | 0.08343 | 0.08993 | 0.08402 | | 0.08818 | 0.09244 | 0.09298 | | 0.09087 | 0.09054 | 0.10282 | 0.09159 | 0.09724 | 0.09774 | 0.09506 | 0.09483 | 0.09521 | 0.09844 | 0.0963 | 0.09636 | 0.1111 | 0.1079 | 0.1003 |
| 0.03 | 0.08091 | 0.0805 | 0.07932 | | 0.08451 | 0.08688 | 0.08509 | | 0.08905 | 0.08881 | 0.09755 | 0.08994 | 0.09679 | 0.09459 | 0.09663 | 0.09677 | 0.09966 | 0.09511 | 0.09363 | 0.09526 | 0.103 | 0.1075 | 0.1035 |
| 0.1 | 0.07916 | 0.08239 | 0.07941 | | 0.0909 | 0.08934 | 0.09185 | | 0.09015 | 0.09026 | 0.10074 | 0.09125 | 0.09921 | 0.09476 | 0.09589 | 0.09534 | 0.0963 | 0.09738 | 0.09643 | 0.09846 | 0.1023 | 0.1011 | 0.1018 |
| 0.3 | 0.08166 | 0.0821 | 0.08222 | | 0.09419 | 0.0911 | 0.09218 | | 0.10039 | 0.10344 | 0.10227 | 0.09614 | 0.10051 | 0.09864 | 0.10078 | 0.10271 | 0.10121 | 0.10223 | 0.102 | 0.1012 | 0.1042 | 0.1038 | 0.1082 |
| 1 | 0.07944 | 0.08142 | 0.07978 | | 0.09762 | 0.09764 | 0.10414 | | 0.10816 | 0.11045 | 0.11651 | 0.10965 | 0.11984 | 0.1151 | 0.11731 | 0.10822 | 0.11639 | 0.11487 | 0.11269 | 0.1142 | 0.1098 | 0.1111 | 0.1109 |
| 5 | 0.06677 | 0.0645 | 0.06725 | | 0.08812 | 0.0839 | 0.08839 | | 0.10233 | 0.10509 | 0.12488 | 0.10008 | 0.11926 | 0.11903 | 0.12575 | 0.12661 | 0.12799 | 0.12677 | 0.12821 | 0.12447 | 0.1365 | 0.1363 | 0.1386 |
| 15 | 0.0597 | 0.05767 | 0.05762 | | 0.07401 | 0.0744 | 0.07536 | | 0.08799 | 0.08659 | 0.1018 | 0.08971 | 0.09857 | 0.09571 | 0.10871 | 0.10926 | 0.10824 | 0.1186 | 0.108 | 0.11901 | 0.1454 | 0.1514 | |
| 50 | 0.0576 | 0.05687 | | | 0.05949 | 0.05765 | | | 0.07021 | 0.06894 | 0.07877 | 0.07538 | | | 0.08711 | 0.08509 | 0.10162 | 0.11564 | | | 0.1452 | 0.136 | |

[a]In the experimental dataset, the data values were fluorescence anisotropy measurements.

[b]In the experimental dataset, the effector was the protein UNG2.

[c]In the experimental dataset, the substrate was the protein PCNA.

Table A2. Starting Values for Parameters During Modeling with Equations 1 and 6

| Parameter | Brain-Cousens Model | | | | | Cedergreen Model | | | |
|---|---|---|---|---|---|---|---|---|---|
| | Raw Dataset | Baseline Subtracted | Normalized & Scaled | Raw Dataset | Baseline Subtracted | Normalized & Scaled Substrate Concentration (µM) | | | |
| | All Substrate Concentrations | | | All Substrate Concentrations | | 0.25 | 0.5 | 1 | 3 |
| a | n/a | n/a | n/a | 0.2 | 0.5 | 0.5 | 0.5 | 0.5 | 0.5 |
| b | 1.5 | 1.5 | 1.7 | 1.5 | 1.5 | 2.0 | 1.3 | 1.3 | 1.3 |
| c | 0.0572[a] | 0[a] | 0[a] | 0.0572[a] | 0[a] | 0[a] | 0[a] | 0[a] | 0[a] |
| d | 0.08 | 0.03 | 100[b] | 0.08 | 0.03 | 100[b] | 100[b] | 100[b] | 100[b] |
| e | 3 | 3 | 3 | 3 | 3 | 3.5 | 8 | 8 | 8 |
| f | 0.01 | 0.01 | 1 | 0.02 | 0.02 | 5 | 100 | 100 | 100 |

[a] parameter $c$ was fixed
[b] parameter $d$ was fixed

Note that successfully chosen starting values only sometimes follow logical trends, highlighting the importance of trial-and-error with starting values and iteratively checking goodness-of-fit of the curves that result from the calculated parameters. The convergence algorithm can be very sensitive to small changes in starting values that unexpectedly lead to different parameter estimates.

**Brain-Cousens SAS Code** ; note that in SAS, natural log "ln" is written as "log"

```
/* Input data for each dataset */
data data_0_25uM;
    input x y gew;
    datalines;
/* Insert data here in column format of x y gew for the 0.25 uM dataset*/
/* Note that gew is equal to 1 divided by standard dev of the replicates*/
;

data data_0_50uM;
    input x y gew;
    datalines;
/*Insert data in the format of x y gew for the 0.50 uM dataset*/
;

data data_1uM;
    input x y gew;
    datalines;
/*Insert data in the format of x y gew for the 1 uM dataset*/
;

data data_3uM;
    input x y gew;
    datalines;
/*Insert data in the format of x y gew for the 3 uM dataset*/
;

data data_10uM;
    input x y gew;
    datalines;
```

```
/*Insert data in the format of x y gew for the 10 uM dataset*/
;

%macro analyze_data(dataset_name);

/* Brain&Cousens for e estimation; c is fixed */
ods output ParameterEstimates=Estimates_Model1;
Proc nlmixed data=&dataset_name;
parms
/* starting values */
b=1.5 d=0.08 e=3 f=0.01;
/* c is fixed */
c=0.0572;
bounds e > 0 ;
eta=(c+(((d-c)+(f*x))/(1+exp(b*log(x/e)))));
model y ~ normal(eta, s2/gew);
run;
ods output close;

/* Retrieve parameter estimates from Model 1 */
proc sql noprint;
    select Estimate into :b_estimated from Estimates_Model1 where
Parameter='b';
    select Estimate into :d_estimated from Estimates_Model1 where
Parameter='d';
    select Estimate into :e_estimated from Estimates_Model1 where
Parameter='e';
    select Estimate into :f_estimated from Estimates_Model1 where
Parameter='f';
quit;

/* Brain&Cousens reparameterized for ED50 (K=50) estimation; c is fixed */
ods output ParameterEstimates=Estimates_Model2;
Proc nlmixed data=&dataset_name;
parms
/* starting values */
ed=30;
/* K is fixed to 50 for ED50 estimation; can be changed to any EDK value
estimation */
b=&b_estimated; d=&d_estimated; f=&f_estimated; K=50; c=0.0572;
e=&e_estimated;
bounds ed > 0 ;
eta=(c+(((d-c)+(f*x))/(1+(((K/(100-K))+((100/(100-K))*(f*ED/(d-c))))*exp(b*l
og(x/ED))))));
model y ~ normal(eta, s2/gew);
run;
ods output close;

/* Retrieve parameter estimates from Model 2 */
proc sql noprint;
    select Estimate into :ed_estimated from Estimates_Model2 where
Parameter='ed';
quit;

/* Brain&Cousens reparameterized for M and ymax estimation */
ods output ParameterEstimates=Estimates_Model3;
```

```sas
Proc nlmixed data=&dataset_name;
parms
M=4;
b=&b_estimated; d=&d_estimated; f=&f_estimated; c=0.0572; e=&e_estimated;
ed=&ed_estimated;
bounds M > 0 ;
eta=(c+((d-c)+(f*x))/(1+(f*M/(((d-c)*b)-f*M*(1-b)))*exp(b*log(x/M))));
ymax = (c+((d-c)+(f*M))/(1+(f*M/(((d-c)*b)-f*M*(1-b)))*exp(b*log(M/M))));
model y ~ normal(eta, s2/gew);
estimate 'ymax' ymax;
estimate 'ymax%' ((ymax*100)/d);
run;
ods output close;

/* Retrieve parameter estimates from Model 3 */
proc sql noprint;
    select Estimate into :m_estimated from Estimates_Model3 where
Parameter='M';
    select Estimate into :ymax_est from Estimates_Model3 where
Parameter='ymax';
    select Estimate into :ymax_percent_est from Estimates_Model3 where
Parameter='ymax_percent';
quit;

/* Brain&Cousens reparameterized for LDS estimation*/
ods output ParameterEstimates=Estimates_Model4;
Proc nlmixed data=&dataset_name;
parms
LDS=10;
b=&b_estimated; d=&d_estimated; f=&f_estimated; c=0.0572; e=&e_estimated;
ed=&ed_estimated; M=&m_estimated;
bounds lds > 0 ;
eta=(c+(((d-c)+(f*x))/(1+(((f*LDS/(d-c)))*exp(b*log(x/LDS))))));
model y ~ normal(eta, s2/gew);
run;
ods output close;

proc sql noprint;
    select Estimate into :LDS from Estimates_Model4 where Parameter='LDS';
quit;

%mend analyze_data;

%analyze_data(data_0_25uM);
%analyze_data(data_0_50uM);
%analyze_data(data_1uM);
%analyze_data(data_3uM);
%analyze_data(data_10uM);
```

**Cedergreen SAS Code** ; note that in SAS, natural log "ln" is written as "log"

```
/* Input data for each dataset */
data data_0_25uM;
    input x y gew;
    datalines;
/* Insert data in the format of x y gew for the 0.25 uM dataset*/
;

data data_0_50uM;
    input x y gew;
    datalines;
/*Insert data in the format of x y gew for the 0.50 uM dataset*/
;

data data_1uM;
    input x y gew;
    datalines;
/*Insert data in the format of x y gew for the 1 uM dataset*/
;

data data_3uM;
    input x y gew;
    datalines;
/*Insert data in the format of x y gew for the 3 uM dataset*/
;

data data_10uM;
    input x y gew;
    datalines;
/*Insert data in the format of x y gew for the 10 uM dataset*/
;

%macro analyze_data(dataset_name);

/*Cedergreen e estimation*/
ods output ParameterEstimates=Estimates_Model1;
Proc nlmixed data=&dataset_name;
parms
/* starting values */
b=1.5 d=0.08 e=3 f=0.02 a=0.2;
/* c is fixed */
c=0.0572;
bounds e > 0 ;
eta = (c+((d-c)+(f*exp(-1/(x**a))))/(1+exp(b*log(x/e))));
model y ~ normal(eta, s2/gew);
run;
ods output close;

/* Retrieve parameter estimates from Model 1 */
proc sql noprint;
    select Estimate into :b_estimated from Estimates_Model1 where Parameter='b';
    select Estimate into :d_estimated from Estimates_Model1 where Parameter='d';
```

```sas
    select Estimate into :e_estimated from Estimates_Model1 where
Parameter='e';
    select Estimate into :f_estimated from Estimates_Model1 where
Parameter='f';
    select Estimate into :a_estimated from Estimates_Model1 where
Parameter='a';
quit;

/*Cedergreen reparameterized for ED50 (K=50) estimation; c is fixed*/
ods output ParameterEstimates=Estimates_Model2;
Proc nlmixed data=&dataset_name;
parms
/* starting values */
ed=30;
/* K is fixed to 50 for ED50 estimation; can be changed to any EDK value
estimation */
b=&b_estimated; d=&d_estimated; f=&f_estimated; K=50; c=0.0572;
e=&e_estimated; a=&a_estimated;
bounds ED > 0 ;
eta = (c+((((((100-K)/100)-(1/(1+exp(b*log(ED/e)))))**(-1))*(((-c+(f*exp(-1
/(ED**a))))/(1+exp(b*log(ED/e))))+(c*(100-K)/100)))-c)+(f*exp(-1/(x**a))))/(
1+exp(b*log(x/e))));
model y ~ normal(eta, s2/gew);
run;
ods output close;

/* Retrieve parameter estimates from Model 2 */
proc sql noprint;
    select Estimate into :ed_estimated from Estimates_Model2 where
Parameter='ed';
quit;

/* Cedergreen reparameterized for M and ymax estimation */
ods output ParameterEstimates=Estimates_Model3;
Proc nlmixed data=&dataset_name;
parms
M=1;
b=&b_estimated; d=&d_estimated; f=&f_estimated; c=0.0572; e=&e_estimated;
ed=&ed_estimated; a=&a_estimated;
bounds M > 0 ;
eta = (c+((d-c)+(((((exp(-1/(M**a))*(a*(M**(-a-1)))*(1+exp(b*log(M/e))))-(ex
p(-1/(M**a))*exp(b*log(M/e))*(b/M)))**(-1))*((d-c)*exp(b*log(M/e))*(b/M)))*e
xp(-1/(x**a))))/(1+exp(b*log(x/e))));
ymax = (c+((d-c)+(((((exp(-1/(M**a))*(a*(M**(-a-1)))*(1+exp(b*log(M/e))))-(e
xp(-1/(M**a))*exp(b*log(M/e))*(b/M)))**(-1))*((d-c)*exp(b*log(M/e))*(b/M)))*
exp(-1/(M**a))))/(1+exp(b*log(M/e))));
model y ~ normal(eta, s2/gew);
estimate 'ymax' ymax;
estimate 'ymax%' ((ymax*100)/d);
run;
ods output close;

/* Retrieve parameter estimates from Model 3 */
proc sql noprint;
    select Estimate into :m_estimated from Estimates_Model3 where
Parameter='M';
```

```sas
    select Estimate into :ymax_est from Estimates_Model3 where Parameter='ymax';
    select Estimate into :ymax_percent_est from Estimates_Model3 where Parameter='ymax_percent';
quit;

/* Cedergreen reparameterized for LDS estimation*/
ods output ParameterEstimates=Estimates_Model4;
Proc nlmixed data=&dataset_name;
parms
LDS=10;
b=&b_estimated; d=&d_estimated; f=&f_estimated; c=0.0572; e=&e_estimated;
ed=&ed_estimated; M=&m_estimated; a=&a_estimated;
bounds LDS > 0 ;
eta=(c+(((((1-(1/(1+exp(b*log(LDS/e)))))**(-1))*(((-c+(f*exp(-1/(LDS**a))))/
(1+exp(b*log(LDS/e))))+c))-c)+(f*exp(-1/(x**a))))/(1+exp(b*log(x/e))));
model y ~ normal(eta, s2/gew);
run;
ods output close;

proc sql noprint;
    select Estimate into :LDS from Estimates_Model4 where Parameter='LDS';
quit;

%mend analyze_data;

%analyze_data(data_0_25uM);
%analyze_data(data_0_50uM);
%analyze_data(data_1uM);
%analyze_data(data_3uM);
%analyze_data(data_10uM);
```

## RStudio code for calculating $R^2$ and Adjusted $R^2$

```r
# Define function to calculate R-squared and adjusted R-squared
r_and_adjusted_r_squared <- function(y_actual, y_predicted) {
  n <- length(y_actual)
  k <- 1 # Number of independent variables (here, it's only 1)

  sse <- sum((y_actual - y_predicted)^2)
  sst <- sum((y_actual - mean(y_actual))^2)

  r_squared <- 1 - (sse/sst)
  adj_r_squared <- 1 - ((1 - r_squared) * (n - 1) / (n - k - 1))

  return(list(r_squared = r_squared, adj_r_squared = adj_r_squared))
}

# Your predicted y-value for a given x
y_predicted <- c(0.08150493, 0.081514244, 0.081539694, 0.081538375,
0.080542747, 0.065994835, 0.058572196, 0.057362795)

y_actual <- list(
  c(0.08343, 0.08993, 0.08402),    # 0.01
  c(0.08091, 0.0805, 0.07932),     # 0.03
  c(0.07916, 0.08239, 0.07941),    # 0.1
  c(0.08166, 0.0821, 0.08222),     # 0.3
  c(0.07944, 0.08142, 0.07978),    # 1
  c(0.06677, 0.0645, 0.06725),     # 5
  c(0.0597, 0.05767, 0.05762),     # 15
  c(0.0576, 0.05687)               # 50
)

# Combine all actual y-values into one vector
y_actual_combined <- unlist(y_actual)

# Calculate how many times each predicted y-value should be repeated
repeats <- sapply(y_actual, length)

# Repeat each predicted y-value according to the repeats vector
y_predicted_combined <- rep(y_predicted, repeats)

# Calculate R-squared and adjusted R-squared values for the entire dataset
r_and_adj_r_squared_values <- r_and_adjusted_r_squared(y_actual_combined,
y_predicted_combined)

# Output the R-squared and adjusted R-squared values
cat("R-squared value:", r_and_adj_r_squared_values$r_squared, "\n")
cat("Adjusted R-squared value:", r_and_adj_r_squared_values$adj_r_squared,
"\n")
```